# Characteristics of Precession Electron Diffraction Intensities from Dynamical Simulations

W. Sinkler and L. D. Marks†

UOP LLC, 50 E. Algonquin Rd., Des Plaines IL U. S. A 60017-5017

†Northwestern University, Department of Materials Science and Engineering, Evanston IL U. S. A. 60208-3108

**Abstract.** Precession Electron Diffraction (PED) offers a number of advantages for crystal structure analysis and solving unknown structures using electron diffraction. The current article uses many-beam simulations of PED intensities, in combination with model structures, to arrive at a better understanding of how PED differs from standard unprecessed electron diffraction. It is shown that precession reduces the chaotic oscillatory behavior of electron diffraction intensities as a function of thickness. An additional characteristic of PED which is revealed by simulations is reduced sensitivity to structure factor phases. This is shown to be a general feature of dynamical intensities collected under conditions in which patterns with multiple incident beam orientations are averaged together. A new and significantly faster method is demonstrated for dynamical calculations of PED intensities, based on using information contained in off-central columns of the scattering matrix.

## Introduction

Precession Electron Diffraction (PED) is a technique for acquiring electron diffraction intensities, invented in 1994 by Vincent and Midgley (Vincent *et al.* 1994), which has recently gained attention and attracted practitioners as the preferred means of acquiring electron diffraction patterns for solving crystal structures, as well as in other diffraction-based applications (Rauch *et al.* 2008). In a PED experiment, the beam incident on the thin crystal is precessed along a cone whose center is the zone axis orientation. The precession intensities thus represent an average over these off-axis incident beam conditions. The individual reflections of the zone axis (out to a maximum spatial frequency which is determined by the cone semi-angle $\varphi$) pass through the surface of the Ewald sphere, and the intensities are integrated over a portion of the reflections' shape factor or relrod. Because of this the intensities better reflect the zone axis Patterson group symmetry, and they are more tolerant of tilt misorientations. In addition, by tilting away from the zone axis, the severity of the multiple-beam dynamical scattering may be decreased somewhat, which can improve the applicability of a kinematical approximation (Own 2005; Sinkler *et al.* 2007; Ciston *et al.* 2008). This latter aspect raises the hope of more reliable, less dynamical electron diffraction intensities for use with crystallographic direct methods for routine structure analysis. This has been borne out in practice by large and increasing number of successful structure analyses which use PED intensities and generally make use of a kinematical or two-beam approximation in treatment of the intensity data as a basis for relating the PED intensities to the underlying structure (Gemmi *et al.* 2002; Gjønnes *et al.* 2004; Dorset *et al.* 2007).

While the successes of PED to date may be considered a breakthrough, it is important to appreciate some of the limitations. The most significant of these is the absence to date of an accurate but simple model for interpreting PED intensities directly in terms of structural information. The two available simplified models are the kinematical approximation and the two-beam model. Variants of the two-beam model have been formulated using different treatments of the precession geometry, for example the Blackman approach, which integrates all reflections over excitation errors from -∞ … ∞, (Blackman 1939; Gjønnes *et al.* 1998) ,or a more exact numerical integration over a range of S determined by a beam's spatial frequency |**g**| (Sinkler *et al.* 2007). Application of one or another of these simplified models is currently essential for extracting crystallographic information from the PED data, for example indications of relative structure factor moduli |U(**g**)|. However, all these models may readily be shown to have quite limited validity, and therefore their use with the data carries a risk that the resulting set of {|U(**g**)|} bears little resemblance to the correct values. The limitations of all models based on kinematical or two-beam approximations have been made clear in previous work using accurate many-beam intensity calculations (multislice or Bloch wave). This demonstration will be briefly reiterated here. In addition, several characteristics of PED, which have been revealed by using dynamical many-beam calculations will be presented. These characteristics may ultimately result in a more robust simple approach to treating PED. This is still a daunting task, however, and it is suggested that ultimately more optimal use of dynamical diffraction for studies of materials with large unit cells may require a rethinking of the experiment.

## Background on Dynamical Calculations

The precession electron diffraction intensity calculations presented here rely on two sets of software. The first is a modification of Northwestern University Multislice and Imaging Software (NUMIS) multislice code, and the second is built around a Bloch-wave formalism to conveniently calculate and store PED intensities. The general approach in both cases is to calculate and store intensities for plane wave incidence along a number of orientations around a precession circuit (typically 1024 settings or more around the circuit). The intensities from each setting are summed and on completion are stored in a single file format, the top row of which contain all the (2-dimensional) h,k values specified for output (listed from left to right by increasing spatial frequency, starting with the 0,0 beam). Each column below an h,k represents the intensity I(**g**,t) of beam **g**=(h,k) for a sequence of thicknesses (generally periodic multiples of the unit cell repeat along the zone axis direction). Each row of the file represents the precession intensities in the zone axis at a fixed thickness. Auxiliary software was written to extract either the I(t) for a selected beam **g**, or the I(**g**) at fixed thickness t (a precession pattern). Another auxiliary software can be used to add all intensities with their inversion-pair and thus halve the computation time for cases of zone axes with inversion symmetry.

Details on the multislice precession simulations can be found in (Own *et al.* 2006). The Bloch wave software is based on a matrix formulation for dynamical intensities similar to that described in (Spence *et al.* 1992; Humphreys 1997), and used the particular arrangement proposed by Allen et al (Allen *et al.* 1998), which is described here briefly since it is used in Section VI below. In that formulation, for a fixed orientation the structure matrix is written as:

$$A = \frac{1}{2K_n}\begin{pmatrix} 2K_n S(\mathbf{h}) & U(\mathbf{h}-\mathbf{g}) & U(\mathbf{h}) & U(\mathbf{h}+\mathbf{g}) & U(2\mathbf{h}) \\ U(\mathbf{g}-\mathbf{h}) & 2K_n S(\mathbf{g}) & U(\mathbf{g}) & U(2\mathbf{g}) & U(\mathbf{g}+\mathbf{h}) \\ U(-\mathbf{h}) & U(-\mathbf{g}) & 0 & U(\mathbf{g}) & U(\mathbf{h}) \\ U(-\mathbf{g}-\mathbf{h}) & U(-2\mathbf{g}) & U(-\mathbf{g}) & 2K_n S(-\mathbf{g}) & U(\mathbf{h}-\mathbf{g}) \\ U(-2\mathbf{h}) & U(-\mathbf{h}-\mathbf{g}) & U(-\mathbf{h}) & U(-\mathbf{h}+\mathbf{g}) & 2K_n S(-\mathbf{h}) \end{pmatrix} \quad (1)$$

where $K_n$ is the length of the electron wave vector, projected onto the sample surface normal (corrected for refraction by the mean inner potential). The U's are structure factors $|U(\mathbf{g})|/2K_n=1/(2\xi_\mathbf{g})$, ($\xi_\mathbf{g}$ is the two-beam extinction distance for reflection **g**) and S(**g**) is the excitation error. In a PED experiment, the excitation error for a ZOLZ reflection **g** is given as (Vincent *et al.* 1994)

$$S(\mathbf{g}) = -\frac{g^2}{2K} + g\varphi\cos\Theta \quad (2)$$

where φ is the precession cone semi-angle, and Θ is the momentary angle between the vector **g** and the center of the Laue circle as the beam moves along the precession circuit. Every column and row of the structure matrix can be thought of as associated with a specific reciprocal lattice vector, ranging in the representation of Eq. 1 from the (**h**,**h**) element in the top left to the (-**h**,-**h**) element in the bottom right. The off-diagonal elements $A_{ij}=U(\mathbf{g}_i-\mathbf{g}_j)/2K$. By solving the eigen equation $\mathbf{AC}=[\gamma]_D\mathbf{C}$ one obtains the eigenvectors of **A** as the columns of **C** and the eigenvalues as the components $\gamma_{jj}$ of the diagonal matrix $[\gamma]_D$. This was done using the Lapack FORTRAN routine zheevd.f (Anderson *et al.* 1999). When **A** is expressed as shown in Eq. 1, with structure factor pairs of +/- **g** arranged about the central element in the central column, the scattered electron wave's Fourier coefficients at thickness t are found as the central column of the scattering matrix **S**, which is obtained as $\mathbf{S}=\mathbf{C}[\Lambda]_D\mathbf{C}^{-1}$, in which $[\Lambda]_D$. is a diagonal matrix with elements $\lambda_{jj}=\exp(2\pi i\gamma_{jj}t)$, and the intensities are the moduli-squared of these central-column elements, for the beams corresponding to the structure factors in the central column of **A**.

For both the multislice and Bloch wave calculations, inelastic scattering is neglected. This is justifiable in the current case, in which the main purpose is to establish general properties of PED intensities, and how they differ from single-orientation transmission electron diffraction patterns. These general distinctions are independent of the absolute accuracy of the calculations and thus for simplicity inelastic scattering was not included. For both multislice and Bloch wave calculations, the neutral atom form factors of Doyle and Turner (Doyle *et al.* 1968) were used.

Many of the questions raised in this work concern issues of how similar two sets of intensities are. For example, in addressing how dynamical or kinematical a set of intensities is we need to compare those intensities quantitatively with a computed kinematical set. For these comparisons, the current work relies on the crystallographic R-factor R1, defined as:

$$R1 = \frac{\sum |I_1(\mathbf{g}) - \alpha I_2(\mathbf{g})|}{\sum_g I_1(\mathbf{g})} \quad (3)$$

For calculating R1, the **g**=0 beam is not included in the sums and $\alpha$ is adjusted to minimize R1. Fig 1 shows a comparison of PED intensities calculated using multislice and Bloch wave approaches for the $(Ga,In)_2SnO_5$ structure (Sinkler *et al.* 1998; Edwards *et al.* 2000), commonly referred to as gallium-indium-tin oxide or GITO. The plot shows the R-factor R1 as a function of thickness for all beams to 1.5 Å$^{-1}$ spatial frequency in this structure. The two computations show excellent agreement, as is revealed by the scatter plot of multislice versus Bloch wave intensities at 1500 Å (Fig. 6b).

## Simplification of diffraction intensities - demonstration using principal component analysis

Principal component analysis (PCA) is a statistical analysis of a set of data 'vectors', or a series of n characteristics of N individuals in a population (Jolliffe 2004). It provides an unbiased assessment of the underlying degree of complexity in the set of n-component vectors. For example, if all vectors are scalar multiples of each other, then 100% of the variation could be accounted for by one component, an n-component unit vector from which all the other vectors could be obtained using appropriate weightings.

PCA was applied to modeled PED I(**g**,t) 'vectors' for the GITO [010] zone axis, for various precession conditions, including the unprocessed ($\varphi$=0) condition (Sinkler *et al.* 2007). The I(**g**,t) vectors were defined as the intensities at increments of the unit cell repeat of 3.16 Å from zero to 948 Å inclusive (n=301). The population consisted of all N=437 unique g-vectors for the zone axis within a resolution circle of 1.5 Å$^{-1}$ radius, i.e. the 'population' is the set of reflections **g**, and the 'vector' for beam **g** is its I(t) curve (set of 301 intensity values from t=0 ... 948 Å) are . The PCA analysis consists of solving the eigenproblem

$$\Sigma\boldsymbol{\alpha} = \lambda\boldsymbol{\alpha} \qquad (4)$$

Where **Σ** is the covariance matrix for the set of N data vectors (the $I_{\mathbf{g}}(t)$) defined as:

$$\sigma_{ij} = \frac{1}{N}\sum_{g} I(\mathbf{g},t_i)\times I(\mathbf{g},t_j) - \bar{I}(t_i)\times\bar{I}(t_j) \qquad (5)$$

Where $\bar{I}(t_i)$ is the average intensity (over all g-vectors) for thickness $t_i$. Solving the eigenproblem Eq. 4 yields the eigenvectors $\boldsymbol{\alpha}$ and eigenvalues $\lambda$. The eigenvectors form a basis set in n dimensions and the largest variation among the N I(t) vectors is with respect to the eigenvector corresponding to the largest $\lambda$. The eigenvalues are in fact the variances of the data with respect to each of the orthonormal eigenvectors (Jolliffe 2004), and the percent of the total variability accounted for by any given eigenvector can thus be expressed as:

$$\sigma_i = 100\% \times \frac{\lambda_i}{\sqrt{\sum_n \lambda_j^2}} \qquad (6)$$

Fig. 2 shows the percent of total variation in the set of $I_{\mathbf{g}}(t)$ accounted for by truncating the expansion of the data vectors at different numbers of eigenvectors, and thus constraining their di-mensionality. This was done for $I_{\mathbf{g}}(t)$ vectors for GITO [010] calculated using multislice for several different values of the cone semi-angle φ at 200 kV. With increasing φ, the dimensionality of the variability among the set of $I_{\mathbf{g}}(t)$ is reduced as shown. The ability to account for all the variation among the different $I_{\mathbf{g}}(t)$ with relatively few basis vectors indicates that using precession simplifies the intensity versus thickness curves, making the intensities more stable and their variation less chaotic. This conclusion is also supported by direct inspection of the I(t) curves, for example that for the GITO (4,0,-1) reflection, which has the largest structure factor in the data. As the precession semi-angle increases, the behavior of I(t) for (4,0,-1) goes from one of rapid oscillations to a single broad maximum and some slow oscillations about a lower value as thickness increases.

## Failure of kinematical and two-beam models

The simplification of intensities versus thickness in PED versus a single-orientation zone axis condition suggests that the intensities themselves may depend on a small number of underlying parameters. This raises the hope that if these parameters and dependences can be identified it might be possible to determine a simple robust model relating PED to the underlying structure factors and experimental conditions. However, in spite of extended attempts, no such model has yet been identified. Parameters on which the PED I(**g**,t) values might at a minimum be expected to depend are the structure factor moduli |U(**g**)| and the spatial frequency |**g**| (the former representing the inherent strength of interaction with the incident beam, and the latter determining the range of excitation errors for a given cone angle φ). For a given experimental geometry (for precession, the semi-angle φ and accelerating voltage) these are the only two quantities which affect the intensity in both the kinematical and 2-beam approximations. Fig. 3 presents I(t) curves for several reflections, calculated using multislice for a [010] zone axis of the zeolite MFI (Olson *et al.* 1981), for 200 kV accelerating voltage and a precession cone semi-angle φ=36 mrad. The reflections plotted were chosen to represent a narrow range of |U(**g**)| and |**g**| values, and yet the curves exhibit a large amount of variation, and no apparent correlation with |U(**g**)| (i.e. the strongest |U(**g**)| is by no means the most intense reflection etc.). This demonstrates that any model for PED in which the intensities depend alone on |U(**g**)| and |**g**| (including kinematical and two-beam approximations) cannot account for PED intensities in any satisfactory way. The simplification of the I(**g**,t) curves obtained by using PED is an important characteristic of PED, and is central to understanding benefits such as somewhat more reproducible sets of relative intensities and the reduction in the amount and severity of false minima in refinements. However, it has not resulted to date in development of an improved simplified model for PED, either by the current authors or in the literature.

## Reduced dependence of PED intensities on structure-factor phases

A central feature of both kinematical and two-beam approximations which contributes greatly to their appeal is their independence of structure-factor phases. Because of this, the structure factor moduli may be deduced directly from experimental data whenever these models are valid, and the additional complexity of a beam's phase ϕ(**g**) need not be considered. By contrast, rigorous dynamical intensity modeling using multislice or Bloch wave approaches requires essentially complete knowledge of the structure, i.e. both structure factor moduli and phases must be known. This raises the question as to whether the phase independence exhibited by the simple kinematical and two-beam models may characterize PED to some limited extent.

In order to address this, Bloch wave PED calculations were carried out using the $(Ga,In)_2SnO_5$ [010] zone axis as a model, with 200 kV accelerating voltage. The U(**g**)'s for the structure matrix were assigned using correct kinematical moduli and random phases (consistent with the centrosymmetric p2 plane group, ZOLZ only reflections). Precession calculations were carried out for six values of φ, increasing from 0 to 60 mrad. In order to evaluate the impact of using random phases, the R-factor R1 was calculated as a function of thickness with respect to precession intensities calculated with the same approach but using the correct phases. The R-factors versus thickness are plotted in Fig 4a. As expected, the R1 starts at zero for small thickness (where kinematical conditions give phase-independence). For any value of φ, there is a steep rise initially to a maximum between about 200-500 Å thickness. However, the extent of the perturbation of intensities due to use of random phases shows a clear decrease as the semi-angle increases, so that the worst disagreement at φ=48 mrad is R1~0.4. Fig. 4b shows a scatter plot of the intensities using random phase versus those using correct phases (omitting the **g**=0 beam), for φ=48 mrad and t=340 Å. As can be seen the agreement is very approximate, but does not exceed some kinematical R-factors for electron diffraction data sets used successfully in structure solutions (Dorset *et al.* 2007). It thus appears feasible in principle to model approximate PED intensities with incomplete structural knowledge, particularly for thicker samples, by using random phases. Unfortunately, the use of random phases does not simplify or speed up the computation. Thus for example a variational approach to determining the structure factor moduli, by fitting them to measured precession intensities in a random phase approximation (for known thickness) does not appear to represent a feasible path to improved analyses.

The observation of a reduced phase dependence of PED intensities raises the question as to how this distinguishing feature results from the PED conditions. Does the reduced phase dependence arise for example as a result of tilting away from the zone axis, or rather is it associated with averaging over many incident beam directions in the PED experiment? In order to address this, the following simulations were conducted: Single-orientation patterns were simulated using 50 different random sets of phases, at orientations with increasing tilt φ from the exact zone axis direction (for this φ-series, a non-special azimuth angle of Θ=47 degrees towards (001) from the reciprocal (100) reflection was kept constant). In addition, at a fixed tilt of φ=36 mrad, for 50 different random phase sets, two series of 3 and 5 patterns respectively were averaged at intervals of ±0.35 degrees centered on the location α=47 degrees from (100). Finally, for 50 different random phase sets, patterns were averaged at 21 positions at constant φ=36 mrad, separated by 1 degree each in Θ, again centered on the same position (φ=36 mrad, Θ=47 degrees). For each beam **g** inside a resolution circle of radius 1.5 $Å^{-1}$, and each thickness t, the average and standard deviation were computed (over the 50 random phase sets). Finally the relative standard deviations were averaged over the entire set of beams **g**, to get a single value as a function of thickness. This averaged relative standard deviation is plotted in Fig. 5a as a function of thickness for the four single orientations, and in Fig. 5b for the four computations at 36 mrad (single orientation and three averaged cases). All single orientations show a rapid increase below about 100 Å followed by a constant plateau with a relative standard deviation close to 1.0 (essentially a standard deviation of the same size, on average, as the beam intensity). By contrast, even the averaging of just three closely-spaced orientations causes a marked decrease in the sensitivity to structure factor phases. Interestingly, this difference is first noticeable at relatively large thickness, i.e. the initial rise at small thickness is the same as for single orientations (although the maximum value decreases as more orientations are averaged together).

The decreased sensitivity of PED intensities to structure factor phases revealed by modeled diffraction using random phases is thus clearly associated with the averaging of orientations. This reduction of phase sensitivity occurs even with quite small variation of the incident beam direction, and a minor reduction in phase sensitivity would thus most likely occur for any non-PED spot pattern, simply due to unavoidable slight convergence of the incident beam. The phase sensitivity decreases as more orientations are added (and more widely varying orientations). However, there is always some fairly considerable sensitivity to structure factor phases for PED, and an approximation based on random phases would be fairly rough. As seen in Fig. 4a, the agreement between a random phase set and the correct structure factor phases for $(Ga,In)_2SnO_5$ corresponds to a worst-case R1 of ~0.4 at large semi-angle φ. This is clearly a fairly rough approximation, but nevertheless better than either kinematical or two-beam approximations for thicknesses beyond a few hundred Å for the case of GITO [010] (Sinkler *et al.* 2007).

## Use of off-central columns of the scattering matrix for PED calculations

Due to Bloch wave periodicity in the reciprocal lattice, the eigenvectors of the structure matrix **A** obtained for an off-axis incident beam condition for which $\mathbf{K}_t=\mathbf{h}$ are related to those for zone axis orientation ($K_t=0$) by:

$$C_{\mathbf{g}1,\mathbf{g}2}(\mathbf{K}_t=\mathbf{h})=C_{\mathbf{g}1+\mathbf{h},\mathbf{g}2}(\mathbf{K}_t=0) \quad (7)$$

This involves a cycling of values within a column of the **C** matrix. Note that the transverse component of the incident beam vector ($\mathbf{K}_t$) is defined here according to convention as the vector from the Laue circle's center to the origin of reciprocal space, i.e. for $\mathbf{K}_t=\mathbf{h}$ the Laue circle's center coincides with the reciprocal lattice vector **-h**. Eq. {7} is only rigorously valid when a quasi infinite set of g-vectors is used, i.e. we can ignore effects at the edges due to swapping rows of **C**. However, it is reasonably well fulfilled if the structure matrix formulation (Eq. 1) extends to large resolution compared to the tilt vector **h**. The eigenvalues of **A** are unaffected by tilting to $\mathbf{K}_t=\mathbf{h}$, except for the addition of a constant factor $|S_\mathbf{h}|=g^2/2K$ (the excitation error of beam **h** referred to the case of $\mathbf{K}_t=0$). The only effect of this constant on the **S** matrix is multiplication by a constant phase factor (see Spence (Spence 1998)). The result is that the off-central columns of the structure matrix **S** contain Fourier coefficients for the scattered electron wave for orientations differing by reciprocal lattice vectors **h** relative to the central column. In particular,

$$|\psi_\mathbf{g}(K_t=\mathbf{h})|=S_{\mathbf{g+h},\mathbf{h}} \quad (8)$$

i.e. the row of S corresponding to reciprocal lattice vector **g+h** and column corresponding to vector **h** in the set of strong beams (see section II).

The availability of beam amplitudes for off-axis incident beam conditions suggests a scheme for improving the efficiency of dynamical PED calculations from the approach described in Section II, which involves on the order of 1000 independent multislice or Bloch wave calculations. In particular, for large projected unit cells, the high density of reciprocal lattice points located along or near the precession circuit may allow the circuit to be approximated by simply using the nearest g-vector orientations, for which intensities can be obtained by solving a single eigen-equation.

This was demonstrated using the $(Ga,In)_2SnO_5$ structure as a model case. In the first approximation, a single zone-axis ($\mathbf{K}_t$=0) Bloch wave calculation was carried out. Beam orientations for a precession circuit of 1024 points at φ=36 mrad were calculated and each orientation was replaced by its nearest g-vector position. In total, 57 orientations were averaged, each weighted by the number of times it was selected as nearest point to the circuit. The agreement between this and the full 1024-point calculation is plotted in Fig. 6 as R1 versus thickness. It is possible to further improve the approximation of the precession circuit by adding more eigen-solutions within the first Brillouin zone, for example at the center of the reciprocal unit cell at $\mathbf{K}_t$ =-(0.5,0.5). The columns of this 2nd S-matrix thus contain diffraction amplitudes for orientations related to the incident beam direction plus any g-vector. Use of these orientations interspersed within those for $\mathbf{K}_t$=0 result in a better approximation of the precession circuit. The additional curves in Fig. 6 shows R1 versus thickness for cases of 2, 4 and 8 eigen-solutions within the reciprocal unit cell. The time required for the 8 eigensolution case was less than 15 minutes on a mid-range laptop (for 3071 beams), plus an additional ~1 minute to read in the eigenvalues and eigenvectors and calculate the intensities for a single thickness. By comparison, the use of 512 discrete eigensolutions (1024 with symmetry) for the PED intensities requires about 12 hours computing time.

The availability of information on off-axis diffraction patterns provides an additional opportunity for insight into the phase-independence arising from averaging over different orientations which was the topic of Section V. Specifically, the phase-independence suggests that the sum or average of the intensities for a specific beam **g** obtained from a single S-matrix by Eq. 8 would result in a quantity which has weakened dependence on the structure factor phases. This was verified by computing **S** for ten different random phase sets at $\mathbf{K}_t$=0, and averaging intensities for all beams **g**, where the average was taken over columns corresponding to tilt ranges inside of limits of ~0 mrad (central column only), 12 mrad (89 columns), 24 mrad (357 columns) and 36 mrad (801 columns). The R-factor R1 was calculated relative to the same calculation wit all correct phases and the average of this (over the ten random phase sets) is plotted as a function of thickness in Fig. 7. As can be seen, the expected reduction of phase-independence found using discrete eigen-solutions at different incident beam conditions is also reproduced in the entries of a single S-matrix.

## Discussion

The present work has explored aspects of PED by using many-beam dynamical simulations. Two central features which have been demonstrated here are

1) the reduction of complexity in the beam intensities versus thickness, demonstrated using principal component analysis

2) the reduced sensitivity of PED intensities to structure factor phases.

Both of these features, and certainly the 2nd, have more general applicability beyond PED, to any case in which diffraction intensities average together even a relatively small amount of variation of the incident beam direction. This was shown here in Figs. 5 and 7, where the averaging of the incident beam direction is either smaller than for PED, or using averaging over tilts corresponding to discrete reciprocal lattice vectors (Fig. 7). Finally, an improved means of com-puting precession using a much smaller number of eigen-solutions has been found, by using off-central columns of the scattering matrix **S**. One underlying assumption in this work is that the best route to understanding the intricacies of PED (and for placing any electron diffraction experiment into a more rigorous relation with the scattering potential) is through the use of many-beam calculations. The argument that this model needs to be modified by adding other phenomena, such as the proposed secondary scattering (Cowley *et al.* 1951; Dorset 2003), does not appear to us to be convincingly demonstrated, and conversely, neither has the inadequacy of the many-beam model. If secondary scattering is occurring, it is notably missing from studies using convergent-beam diffraction (which, with energy filtering can provide highly accurate agreement with many-beam theory (Tsuda *et al.* 1999)). Other problems involve the use of this phenomenon to simulate patterns, for example the presence of adjustable parameters and the entire question of how to combine it in some stable form with many-beam scattering.

The motivation for the current work was not limited to better understanding the properties of the PED intensities, including how well they approximate two-beam or kinematical models, but it was also hoped to develop improved models for relating the PED data to the underlying structure. In this last aspect, the work has fallen short of its goal and it is worth assessing the prospects for success in this regard. While the advantages of PED described in the Introduction are well known in the community of electron crystallographers, the attempt to place PED on a more rigorous footing has not yielded much in the way of concrete useful advances. This suggests that the complexity of the physics relating the underlying structure investigated to the observed intensities is simply so great as to be intractable. In spite of the significant advance that it represents, PED may not fully answer the need for a technique which is convenient, applicable to unknown structures with large unit cells (and beam sensitivity), and yet which also permits rigorous measurement of structure factor moduli.

The availability of g-vector tilted beam amplitudes in the scattering matrix suggests there may be much benefit in modifying the precession experiment to collect a series of diffraction patterns for which $\mathbf{K}_t$ is adjusted sequentially to coincide with a set of g-vectors. Such an experiment would provide a close equivalent of PED by averaging intensities taken at selected orientations. However whatever it may lose in convenience (time to acquire and process the patterns, as well as somewhat less suitability to beam-sensitive materials) it may gain in the enhanced ability to use the much greater accuracy of dynamical calculations as a physical model. The notion of using multiple diffraction patterns acquired with different incident beam conditions has been proposed by other authors in the context of inversion algorithms for directly obtaining the scattering potential (Allen *et al.* 1998; Spence 1998; Allen *et al.* 1999; Spence *et al.* 1999; Allen *et al.* 2008). There is a large range of possibilities as to how such data might most effectively be used. One point of particular interest which has long been recognized and is evident in some of the simulations shown above is the sensitivity of dynamical intensities to the structure factor phases. Therefore, due to the relative speed with which a single eigensolution can be obtained, the availability of several patterns taken at different orientations may provide the basis for a physical test of the suitability of different sets of phases resulting from direct methods. Nevertheless, there are many challenges to overcome in this area. The dynamical calculations are highly sensitive to accuracy of the inputs and intolerant of incompleteness (e.g. limited resolution). Some of the proposed experiments are very far from the simplicity and broad applicability of PED, and real applications to unknown structures (or even large unit cells) are missing. In spite of the many challenges, this type of approach using precession devices (now widely available) for multiple-orientation patterns may be the most promising way to integrate rigorous many-beam theory with electron diffraction experiments for solving unknown structures with large unit cells.

*Acknowledgements.* W. S. would like to acknowledge the support of UOP Honeywell for this work.  Funding for L. D. M is through U. S. Department of Energy on Grant Number DE-FG02-01ER45945/A007.

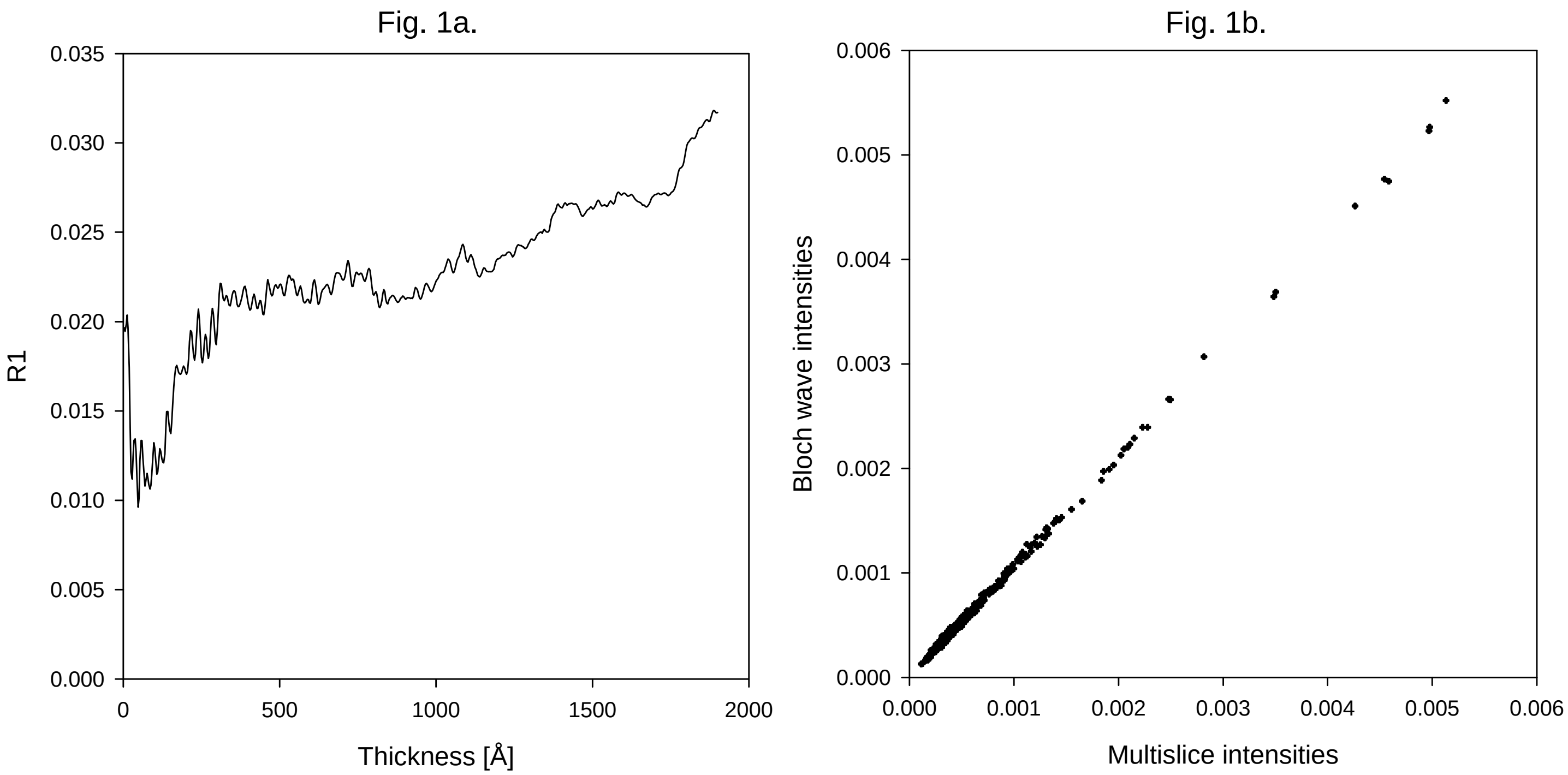

**Fig. 1a. R-factor R1 vs. thickness showing agreement between Bloch wave and multislice precession simulations for $(Ga,In)_2SnO_5$, φ=36 mrad, 200 kV. Conditions for Bloch wave are 3071-beam calculations (all beams below 2.81 Å-1), using 1024 settings around the precession circuit. For multislice the conditions were the same except the potential was sampled to ~10 Å-1 and the repeat distance was 1.58 Å (1/2 of the b-axis length). Fig. 1b shows a scatter plot of intensities at 1500 Å thickness for all beams except g=0.**

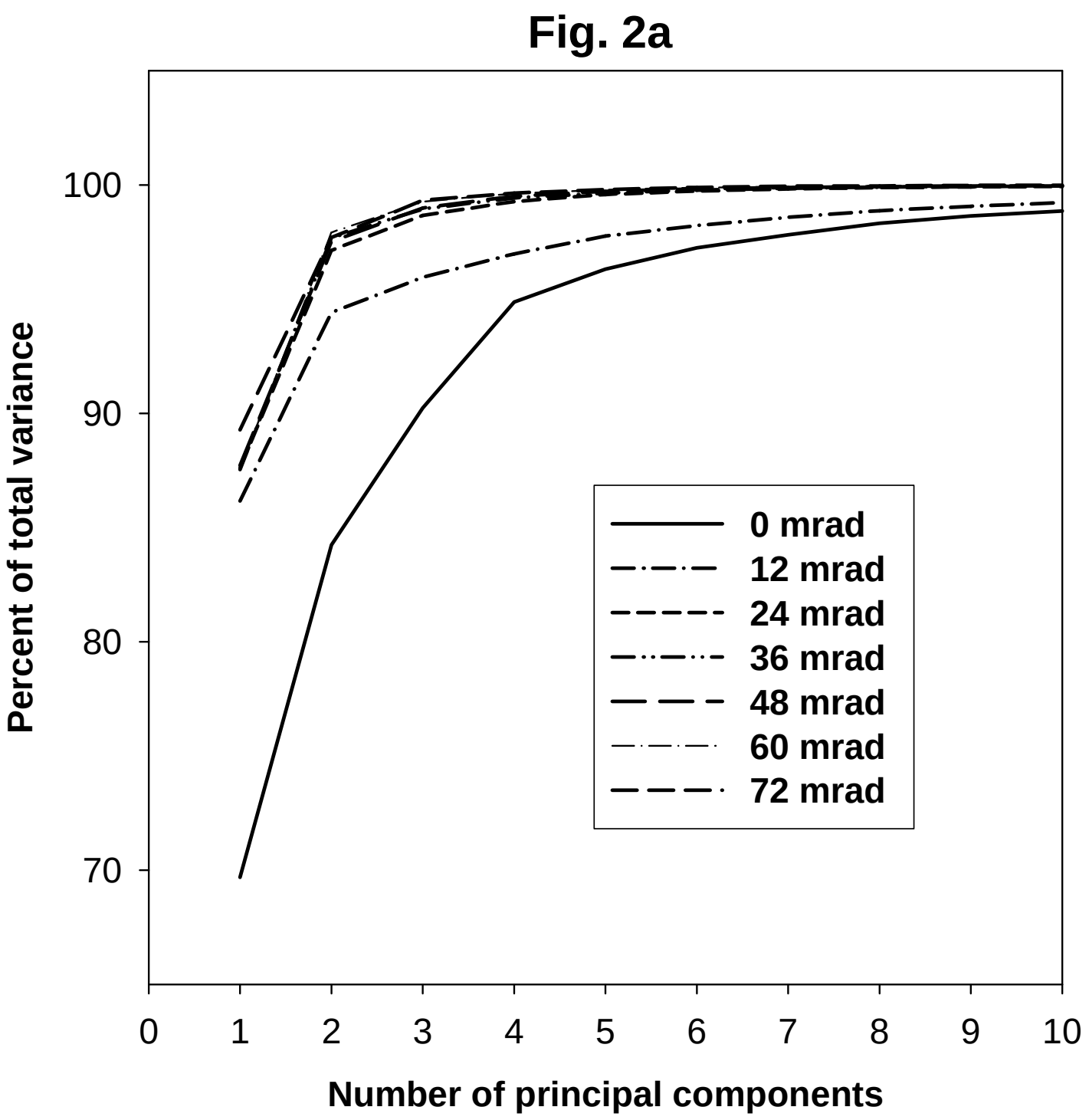

**Fig. 2a. Percent of total variation in set I(g,t) as a function of the number of principal components, for calculations using a range of cone semi-angles φ.**

**Table 1.** Percents of variation in I(g,t) for different numbers of principal components and cone semi-angles φ.

| No. of PC's | φ=0 | φ=12 | φ=24 | φ=36 | φ=48 | φ=60 | φ=72 |
|---|---|---|---|---|---|---|---|
| 1 | 69.7 | 86.2 | 87.6 | 87.5 | 87.7 | 87.5 | 89.3 |
| 2 | 84.2 | 94.4 | 97.1 | 97.7 | 97.5 | 97.9 | 97.7 |
| 3 | 90.2 | 95.9 | 98.7 | 98.9 | 99.0 | 99.3 | 99.3 |
| 4 | 94.9 | 97.0 | 99.3 | 99.4 | 99.5 | 99.6 | 99.6 |
| 5 | 96.3 | 97.8 | 99.6 | 99.6 | 99.7 | 99.8 | 99.8 |

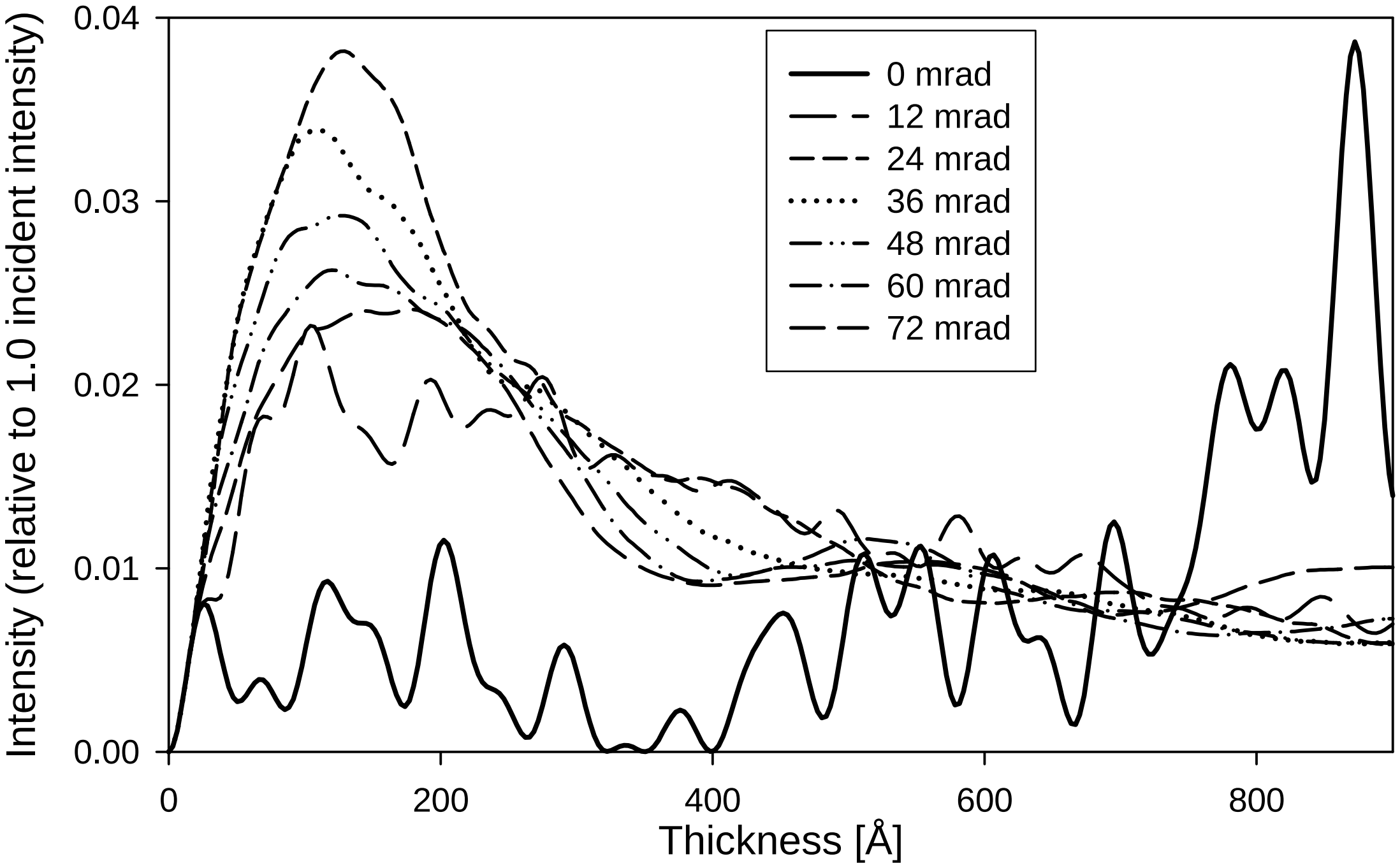

**Fig. 2b. plots of I(t) for GITO (4,0,-1) reflection in the [010] zone axis, from multislice, for increasing precession semi-angle φ. With the introduction of precession, the behavior transitions from one of chaotic oscillation to slower-varying and simpler form, consistent with the statistical simplification determined with principal components analysis.**

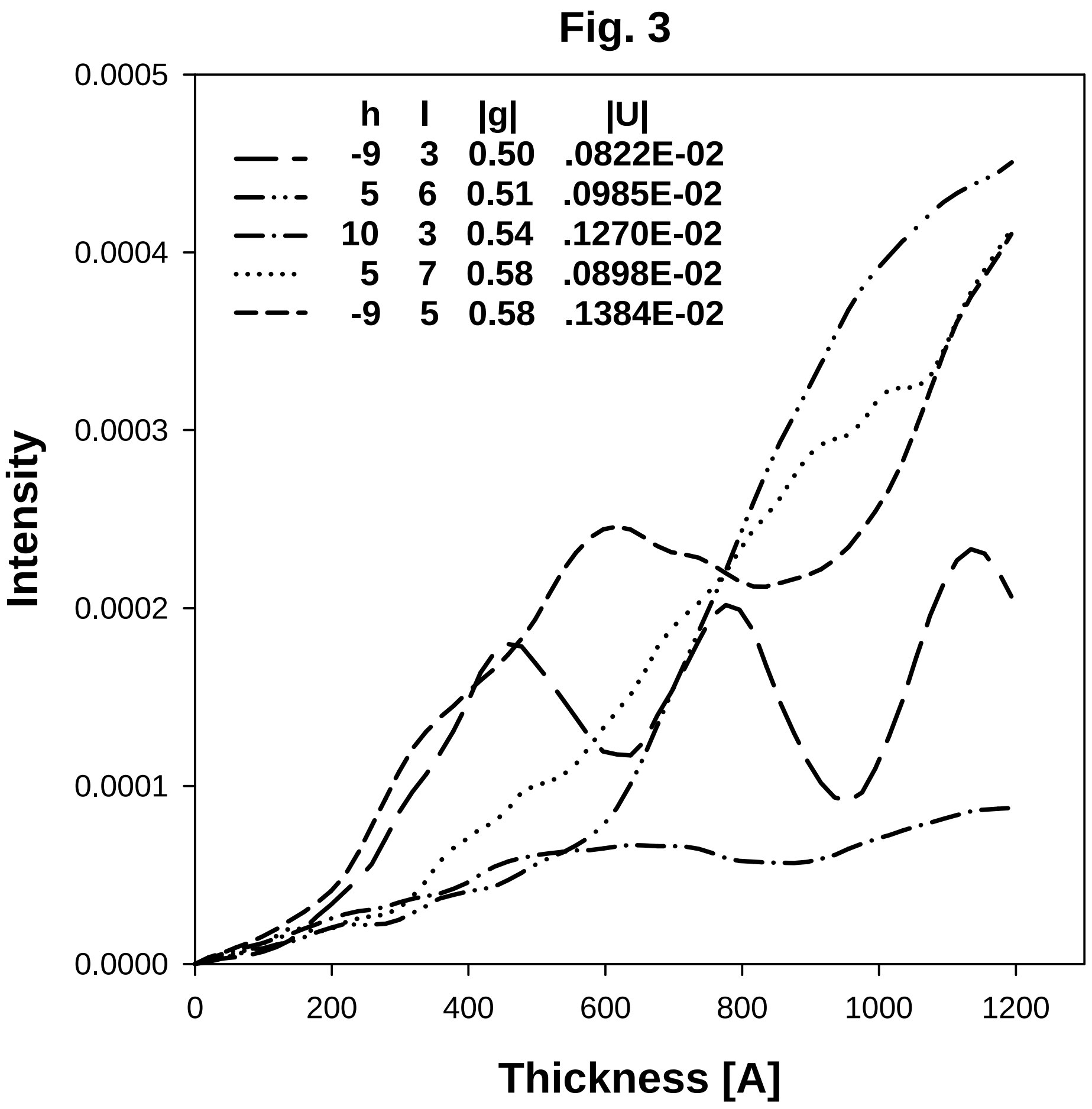

**Fig. 3 I(g,t) plots from multislice calculations for zeolite MFI in [010] zone axis orientation (φ=36 mrad, 200 kV). The reflections plotted were chosen from a narrow range of |g| and |U(g)| values (see legend). The variation among the curves is considerable, and shows no apparent correlation with either |g| or |U(g)|.**

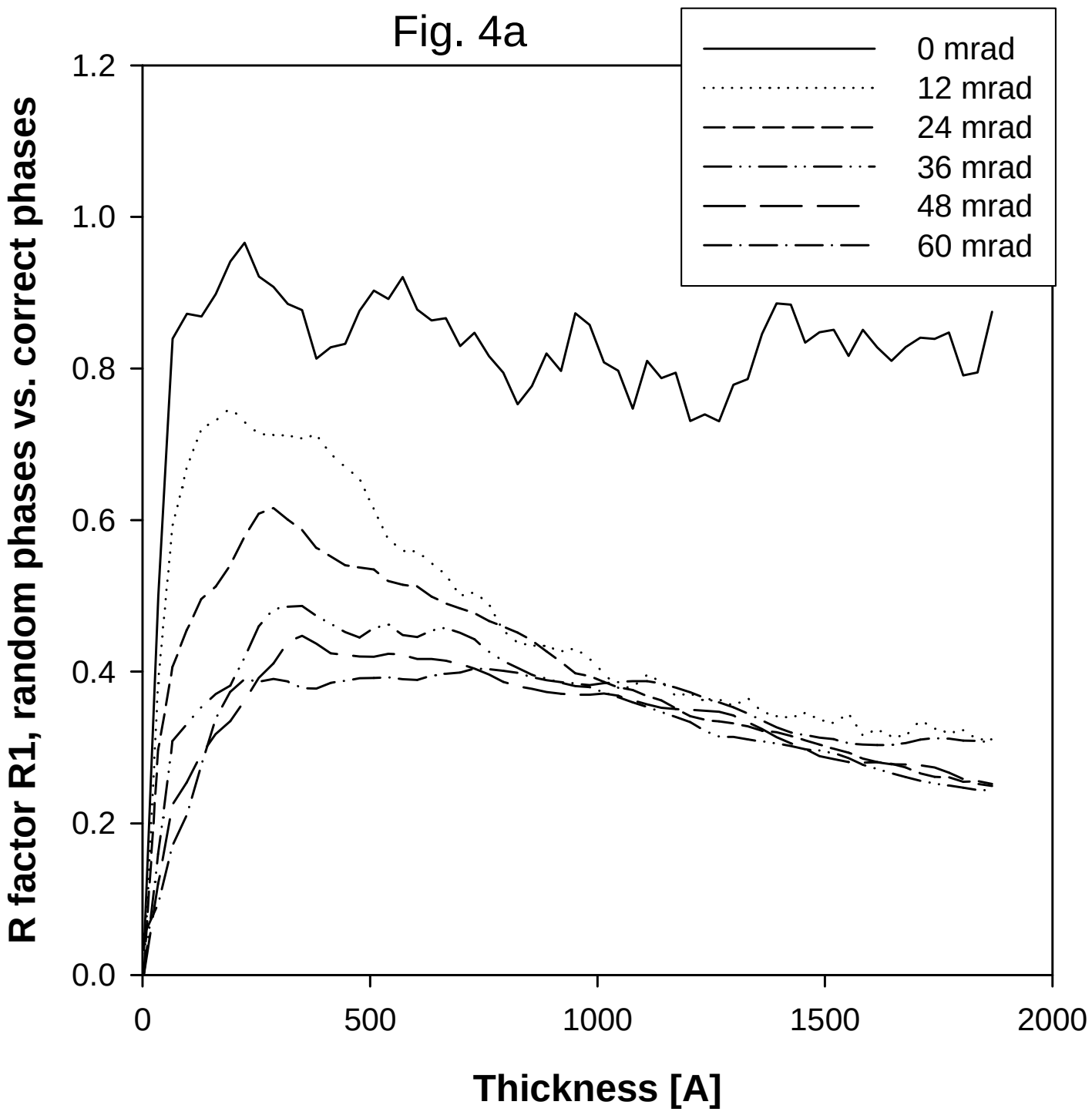

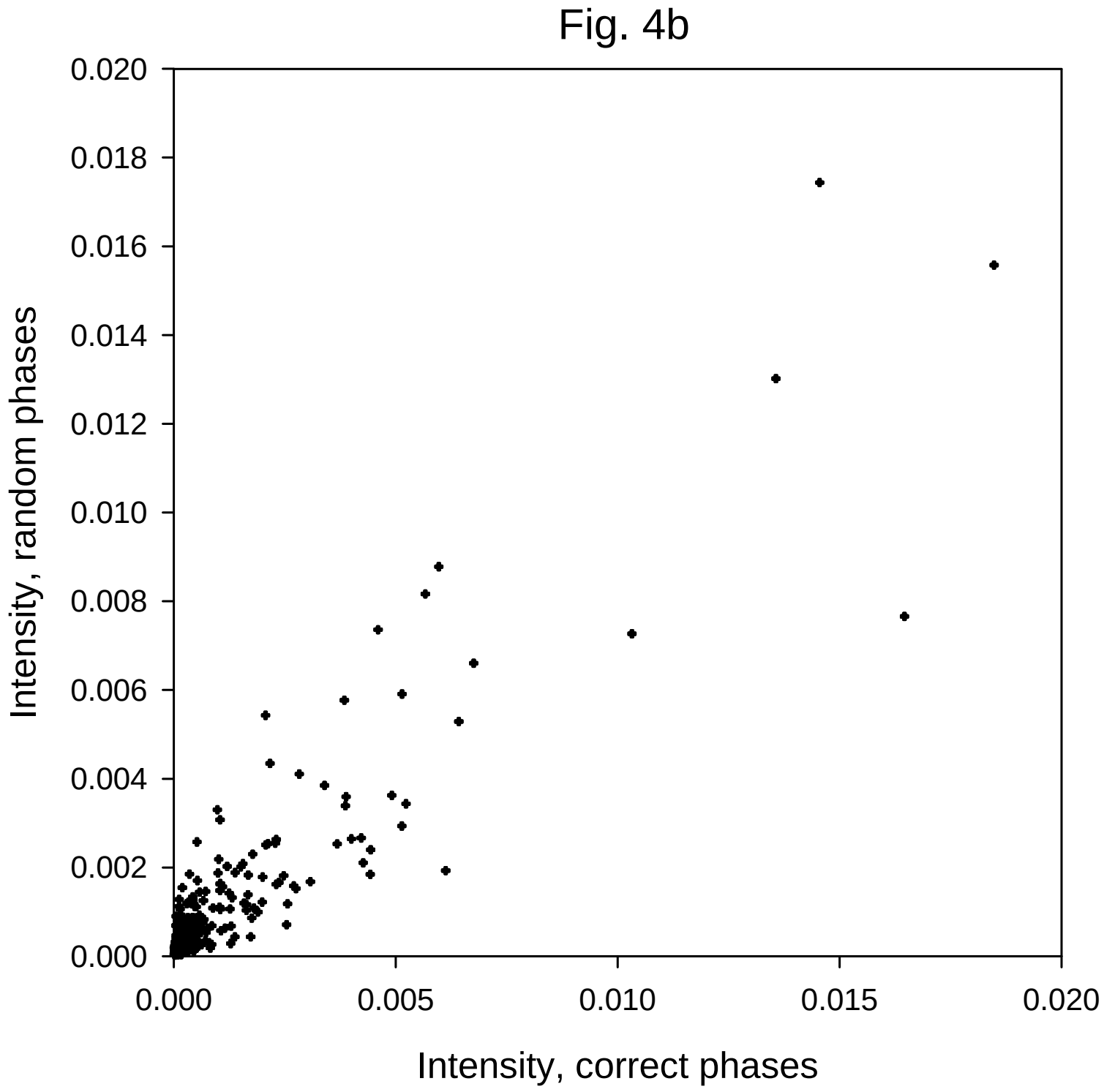

**Fig. 4a Plot of R1 for PED intensities in GITO [010] zone axis, calculated using random phases. The R-factor is calculated relative to the intensities simulated with all correct structure factor phases. Fig. 4b. Illustration of the degree of agreement between intensities simulated using all correct phases and intensities using random phases, for the case of φ=48 mrad, 345 Å thickness.**

Fig. 5a

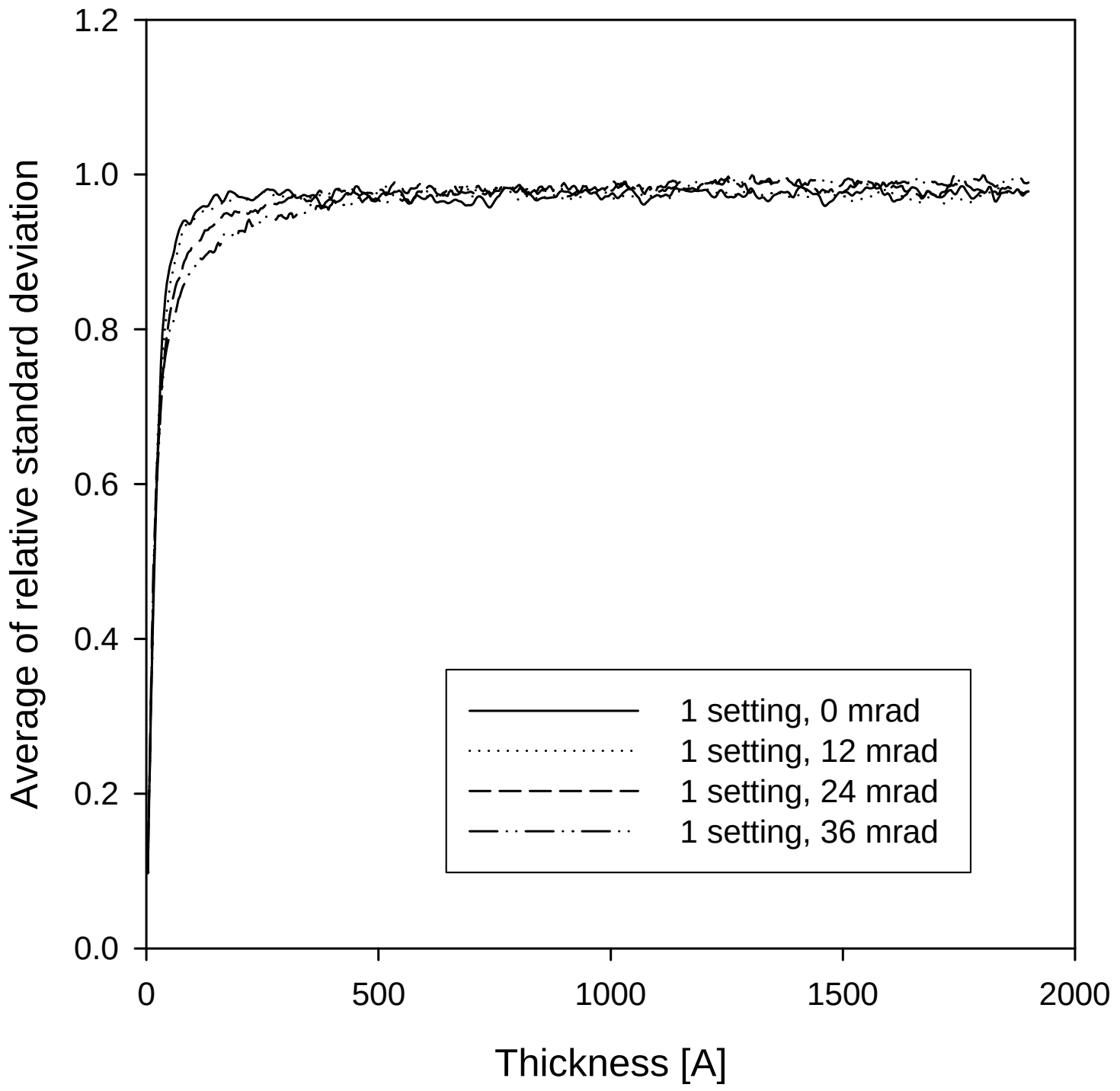

Fig. 5b

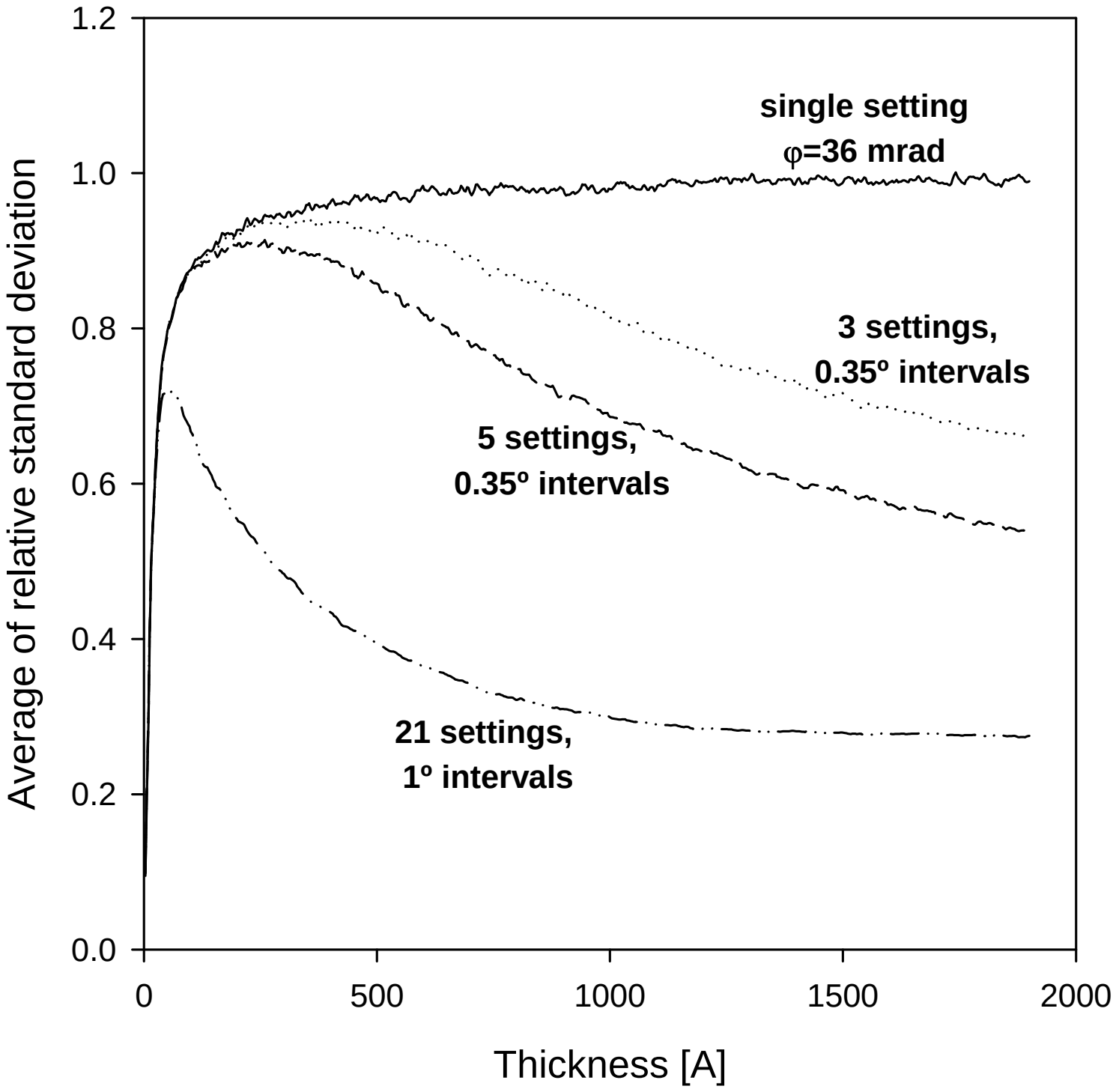

**Fig. 5a. The value of the relative standard deviation, averaged over 437 beams within 1.5 Å$^{-1}$, from Bloch wave calculations of the GITO [010] zone axis for 50 different random phase sets. Four curves are shown at varying φ (with Θ fixed at 47 degrees from (100)). 5b. Calculations using the same 50 random phase sets, this time using averaged incident beam orientations, all with φ=36 mrad (3, 5 and 21 different orientations averaged as indicated). Single setting curve (φ=36 mrad; Θ=47 degrees) is reproduced from Fig. 5a.**

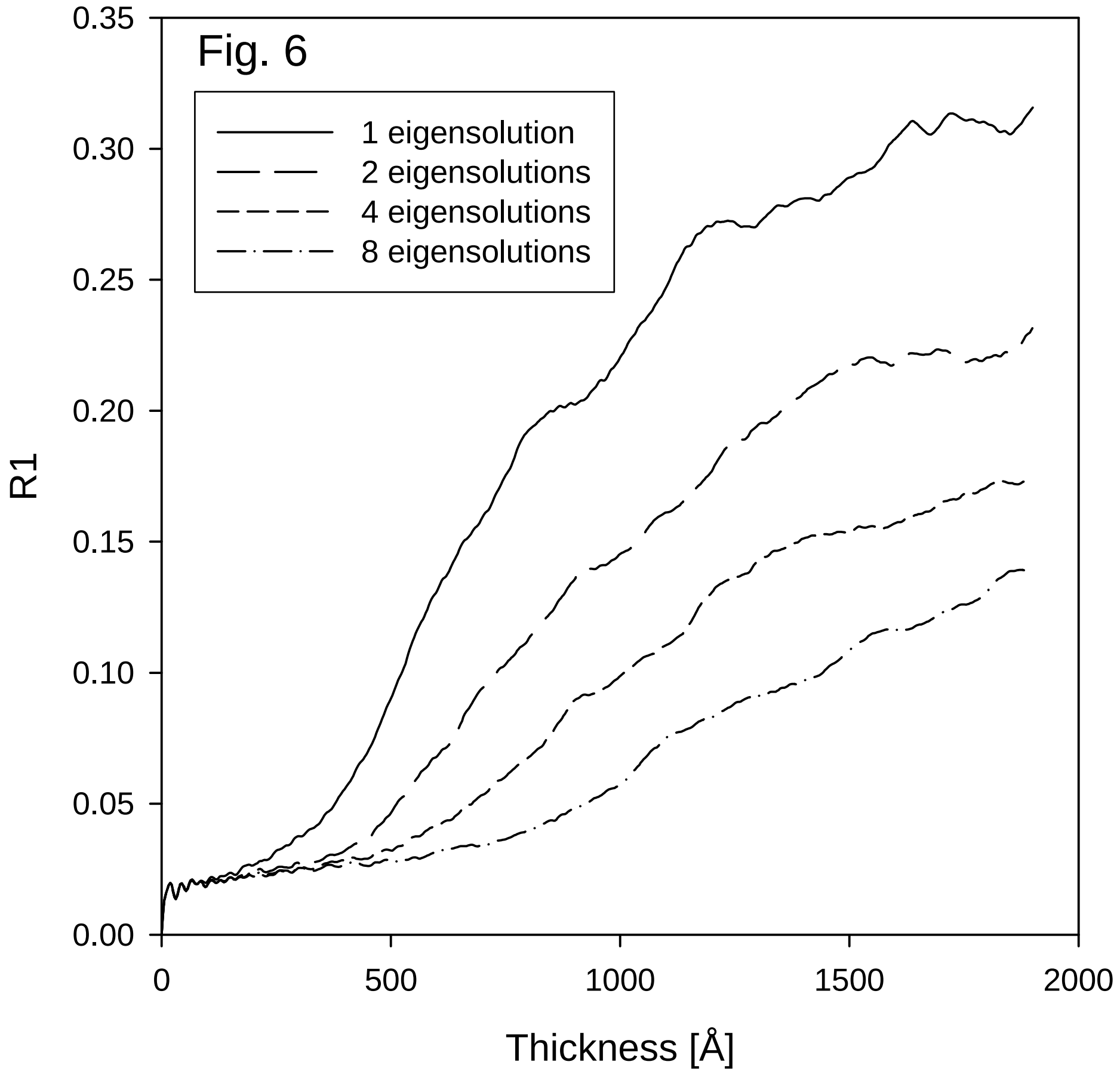

**Fig. 6. Plot of R1 versus thickness for PED Bloch wave calculations utilizing off-central columns of the scattering matrix to obtain intensities at reciprocal lattice tilts. The curves represent different numbers of eigensolutions for tilts within the reciprocal unit cell. As the number increases, the approximation of the precession circuit improves.**

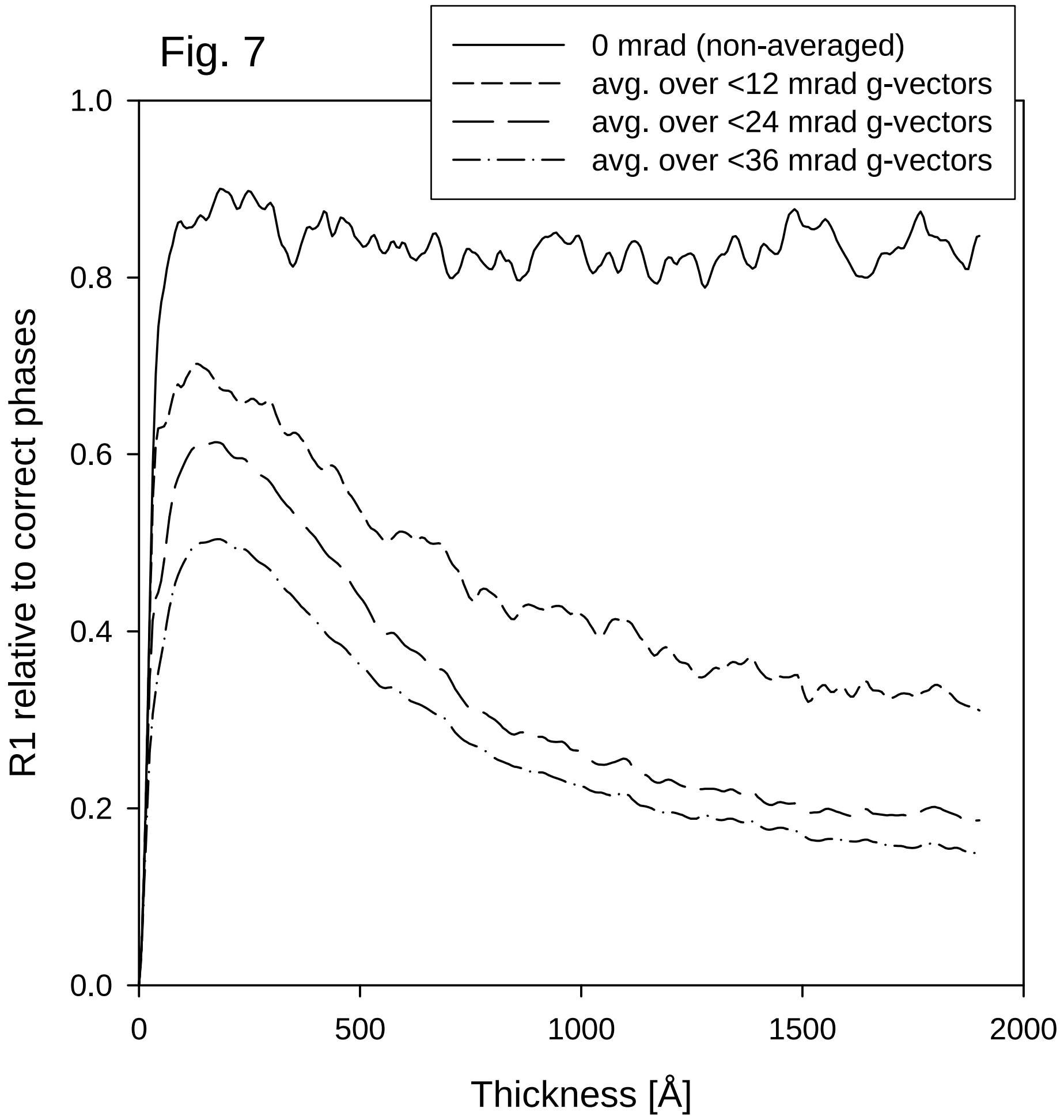

**Fig. 7. Plot R1 for random phases relative to calculation with correct structure factor phases. Single Bloch wave calculations were performed for ten random phase sets for the GITO [010] zone axis. For each of these, intensities were averaged from a number of off-central columns of the scattering matrix to simulate all g-vector tilts located within cones of the indicated semi-angle. The 0, 12, 24 and 36 mrad cases average over 1, 87, 357 and 801 orientations respectively. For each random phase set, R1 was computed with respect to the correct phase set and the average of the ten R1 vs. thickness curves is plotted.**